# Two-Photon Photocurrent in InGaN/GaN Nanowire Intermediate Band Solar Cells

Ross Cheriton[1,2*], Sharif M. Sadaf[1,3,4*], Luc Robichaud[2], Jacob J. Krich[2], Zetian Mi[4,5], Karin Hinzer[2]

[1]Advanced Electronics and Photonics, National Research Council of Canada, 1200 Montreal Rd, Ottawa, ON, Canada, K1A 0R6
[2]Centre for Research in Photonics, University of Ottawa, 25 Templeton St, Ottawa, ON, Canada, K1N 7N9
[3]Centre Energie, Matériaux et Télécommunications, Institut National de la Recherche Scientifique (INRS), 1650 Boulevard Lionel-Boulet, Varennes, QC, Canada, J3X 1S2
[4]Department of Electrical and Computer Engineering, McGill University, 3480 University Street, Montreal, Quebec, Canada, H3A 0E9
[5]Department of Electrical Engineering and Computer Science, University of Michigan, 500 S State St, Ann Arbor, MI, USA, 48109
*these authors contributed equally to this work

**Intermediate band solar cells hold the promise of ultrahigh power conversion efficiencies using a single semiconductor junction. Many current implementations use materials with bandgaps too small to achieve maximum efficiency or use cost-prohibitive substrates. Here we demonstrate a material system for intermediate band solar cells using InGaN/GaN quantum-dot-in-nanowire heterostructures grown directly on silicon to provide a lower cost, large-bandgap intermediate band solar cell platform. We demonstrate sequential two-photon current generation with sub-bandgap photons, the hallmark of intermediate band solar cell operation, through vertically stacked quantum dots in the nanowires. Near-infrared light biasing with an 850 nm laser intensity up to 200 W/cm$^2$ increases the photocurrent above and below the bandgap by up to 19% at 78 K, and 44% at room temperature. The nanostructured III-nitride strategy provides a route towards realistic room temperature intermediate band solar cells while leveraging the cost benefits of silicon substrates.**

## Introduction

Nanowire-based devices are used for highly tunable III-V optoelectronics grown on affordable, lattice-mismatched substrates such as silicon. Using III-nitride materials in a nanowire geometry, devices are developed for use as light emitting diodes[1,2], high electron-mobility transistors[3], photodetectors[4], lasers[5–7], and solar cells[8]. In the case of solar cells, attaining high efficiency at a reasonable cost is crucial for a viable platform. With traditional silicon-based solar cells achieving efficiencies of just over 26%[9], they are approaching their fundamental limiting efficiency of around 32%. While traditional solar cells are subject to the Shockley-Queisser limit[10], intermediate band solar cell (IBSC) concepts increase both current and voltage while still using a single junction[11]. Such designs enable the harvesting of energy from sub-bandgap photons through intermediate states deep inside the semiconductor bandgap that act as steppingstones for photogenerated carriers to reach the conduction band while operating at the higher voltage associated with the large bandgap. IBSCs have the potential to reach ultrahigh efficiencies in excess of 45% (and over 60% with concentration[12]), equivalent to a triple-junction solar cell, without the materials, tunnel junctions, number of layers, and cost associated with multijunction solar cells.

The formation of an intermediate band has been pursued using highly mismatched alloys, quantum dot systems and hyperdoping, as outlined in a review by Okada et al.[12] A common route to produce IBSCs is through InAs (0.35 eV) quantum dot arrays in GaAs (1.4 eV) using established growth processes of the InAs/GaAs quantum dot system[13]. While InAs/GaAs quantum dots have proven useful for tuning the bandgap of individual subcells in the regime of thermionic carrier escape in multijunction solar cells[14], as IBSC candidates they have sub-optimal theoretical power conversion efficiency[11] and suffer from losses through carrier escape out of the quantum dots at room temperature[15]. Ideally, with a 6000 K black body spectrum, the optimal bandgaps for an IBSC should be 1.95 eV and 0.7 eV under full concentration, and 2.4 eV and 0.9 eV under 1 sun illumination[12]; these high bandgaps are unavailable with most III-V semiconductors. The InGaN/GaN material system supports the ideal bandgap combinations for intermediate band operation[16], has a strong absorption coefficient[17], and also benefits from some commercial maturity from light emitting diode technology. The bandgap of InGaN alloys ranges from 3.4 eV to 0.7 eV, spanning the solar spectrum and the ideal intermediate band transition energies between bands. InGaN quantum dots in planar GaN on AlN/sapphire substrates have previously demonstrated the sequential two-photon sub-gap absorption that is the hallmark of intermediate band activity[18].

We improve on those results by using InGaN quantum dots inside GaN nanowires on silicon substrates, showing strongly increased sequential two-photon carrier generation while using an inexpensive substrate and higher indium compositions. The nanowire growth mode removes the need for a lattice-matched substrate and supports vertical stacking of multiple quantum dots without the formation of extended defects[19]. In such a platform, IBSC designs would support higher efficiencies and superior light trapping while retaining the cost advantages of silicon substrates. We demonstrate that the sequential 2-photon subgap photoresponse in our nanowires on silicon is non-thermionic and is significantly stronger than found with ZnTe:O[20], InAs/GaAs[13], GaAs:N[21,22], and even previous InGaN/GaN[18]. We show that this sequential 2-photon photocurrent at room temperature is non-thermionic and that with light bias, the solar cells exhibit 44% increase in photocurrent at room temperature, and 19% increase in subgap photocurrent at 78 K.

## Results
### Figure of merit

A necessary condition for an efficient IBSC is for the intermediate band absorber material to have a high figure of merit $v=E_g\mu\tau\alpha^2/q$, where $\alpha$ is the sub-bandgap absorption coefficient, $\mu$ is the carrier mobility, $\tau$ is the carrier lifetime, $E_g$ is the bandgap, and $q$ is the elementary charge[23,24]. The figure of merit captures the trade-off between increased sub-bandgap absorption and increased carrier recombination due to the introduction of an intermediate band. While an accurate determination of the electron figure of merit requires knowledge of the largely unknown intermediate band to conduction band absorption cross-section, we expect the hole figure of merit for InGaN/GaN systems to be relatively high as a result of the strong bulk material interband absorption (>5x10$^4$ cm$^{-1}$) and large interband transition energies (>2 eV), despite short non-radiative carrier lifetimes[19] (~1 ns) and low hole mobilities (~10 cm$^2$V$^{-1}$s$^{-1}$).

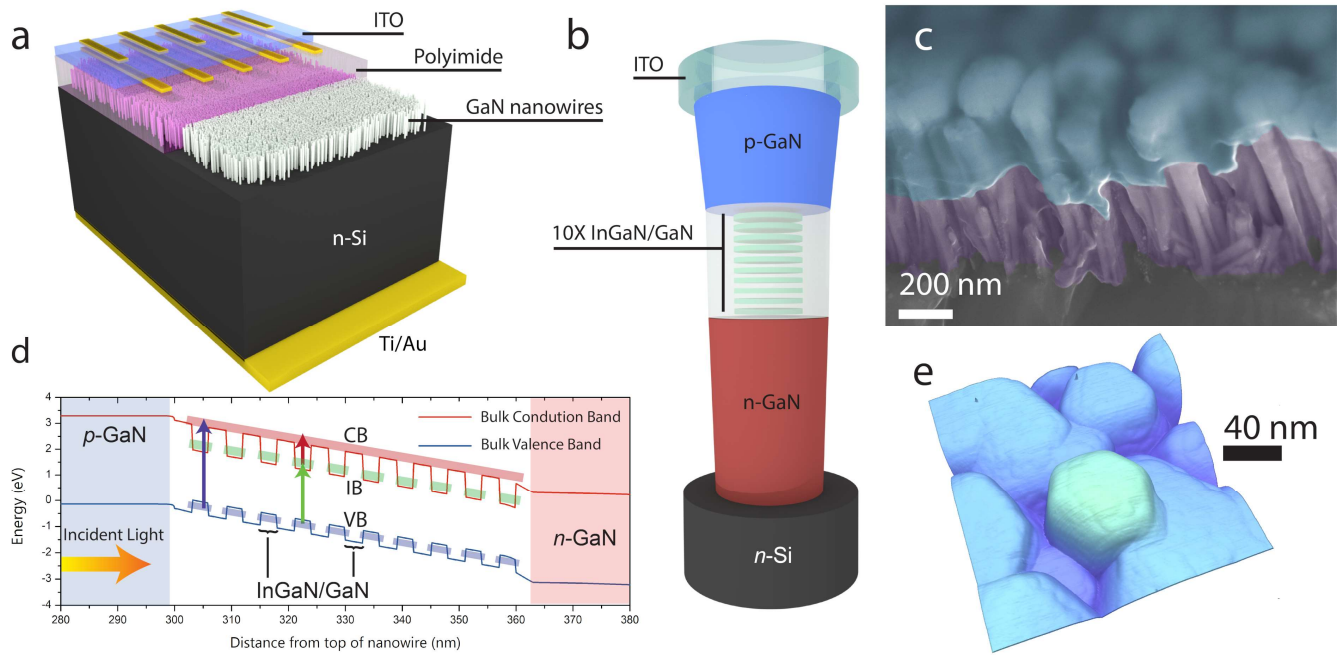

**Figure 1 | Cell design and operation. a** Schematic of the nanowire solar cell (not to scale) depicting the nanowires grown on silicon and capped with gold and indium tin oxide (ITO) contacts. **b** Individual nanowire structure with a cutaway of the active region for subgap photocurrent generation (not to scale). **c** Cross-sectional scanning electron microscope image of the fabricated nanowire solar cell with ITO in blue, nanowires in violet, and silicon in grey. **d** Simulated band diagram (without polarization effects) at short-circuit highlighting the broadband absorption of light through intermediate states. The electric field is introduced by the *p-i-n* architecture. **e** Atomic force microscopy of the top surface of the nanowire ensemble showing the hexagonal shape of the bare nanowires.

**Nanowire design and geometry**

A dense, random ensemble of c-plane nitrogen-face GaN nanowires, each containing ten InGaN quantum dots, was grown by molecular beam epitaxy directly on a silicon substrate, as shown in Figures 1a, 1b, and 1c. The indium composition in the quantum dots varies, ranging up to approximately 40% with variation between and within nanowires. In the nanowires, the quantum dot states are inherently decoupled from the conduction and valence band states as a result of strong carrier confinement in the dots. This decoupling of the quantum dot region is provided by 3 nm barriers of GaN between dots, as shown in the band diagram in Figure 1d. The quantum dots have diameters of about 40 nm and heights of 3 nm. Transmission electron microscopy and growth details of such nanowires have been previously described with the quantum dot indium composition studied in great detail[25]. We imaged the nanowires using scanning electron microscopy and atomic force microscopy to assess their morphology, as shown in Figure 1c and 1e. The bare hexagonal nanowire ensemble is densely packed with an areal density of 10$^{10}$ nanowires per cm$^2$, as shown in Supplementary Fig. 2 with an average nanowire diameter of 89 nm determined statistically in Supplementary Fig. 3. The nanowires are grown without any catalysts and do not rely on substrate patterning techniques that can introduce significant fabrication cost.

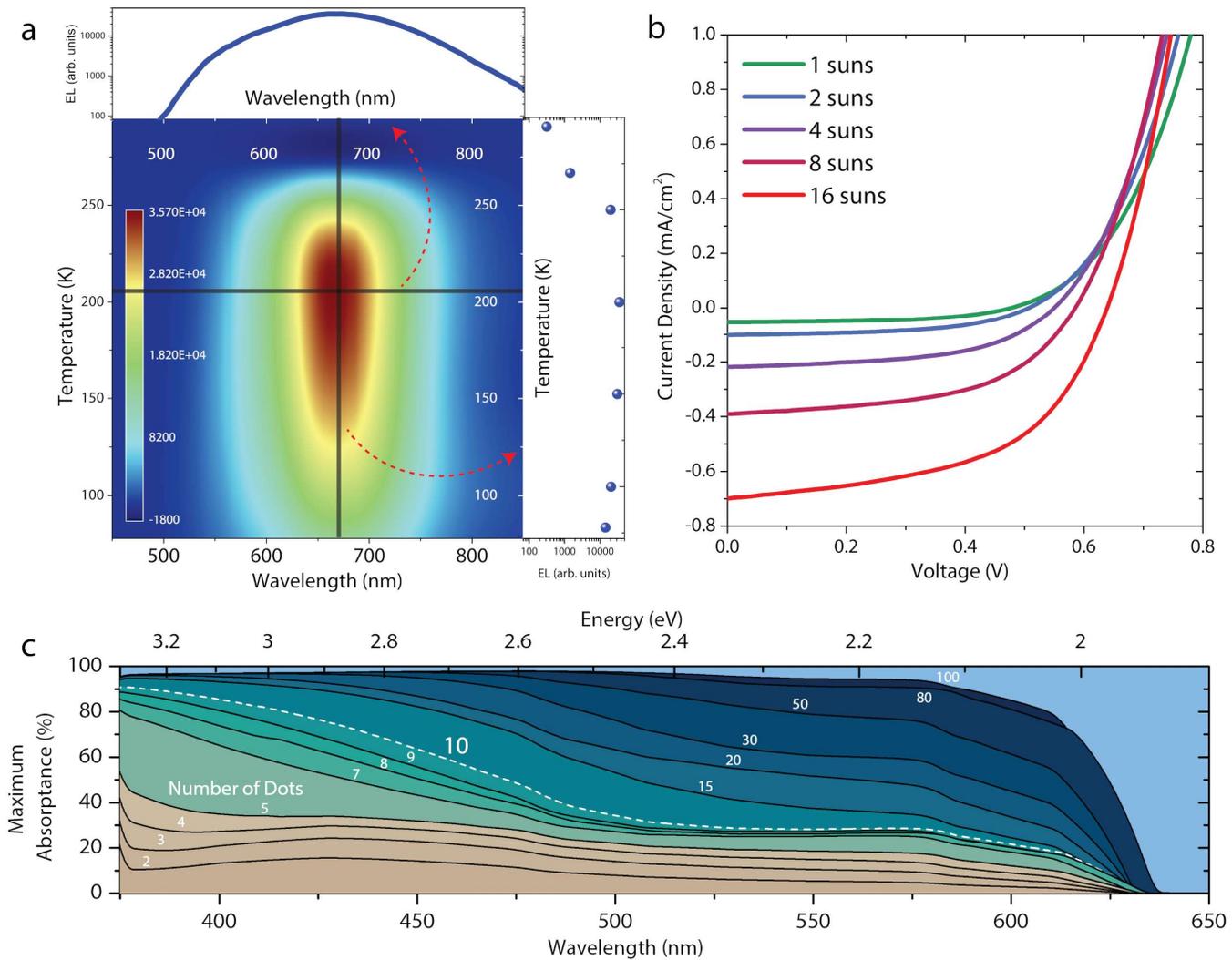

**Figure 2 | Optoelectronic characterization and simulation. a** Electroluminescence (EL) spectra as a function of temperature with corresponding slices through the center wavelength and temperature of maximized signal. **b** Current-voltage characteristics as a function of AM1.5D illumination up to 16 suns concentration. **c** Simulated absorptance for the valence band to intermediate band transition as a function of incident photon wavelength and number of dots.

**Optoelectronic characterization**

    The emissive properties of the quantum dot-in-nanowire solar cells are probed through electroluminescence spectroscopy as a function of temperature. An electrical bias of 4 V is chosen to produce significant current in forward bias through the solar cell, for operation as a light emitting diode. The radiative recombination inside the quantum dots allows determination of the transition energies between the confined electron and hole quantum dot states. Increasing the bias voltage introduces a blueshift of the output spectrum a shown in Supplementary Fig. 4. Electroluminescence spectroscopy at 4 V bias reveals broad emission from the quantum dots from about 550 nm to 750 nm (Figure 2a), indicating a broad distribution of quantum dot state energies. With reduced temperatures, the electroluminescence reaches a peak at around 200 K at 670 nm. The increase in luminescence with temperature down to 200 K is attributed to the reduction in phonon-assisted recombination of carriers inside the quantum dots nanowires. The decrease in electroluminescence below 200 K is a result of the high magnesium dopant activation energy (~0.15 eV)[26] which decreases the *p*-type dopant activation. The electroluminescence spectrum from the nanowire solar cells depends on the indium composition, geometries and dimensions of the quantum dots. The peak of the electroluminescence spectrum is consistent with the radiative recombination from multiple intermediate levels to the valence band. The broad electroluminescence spectral feature indicates that indium composition fractions can be over 40%, corresponding to an effective bulk InGaN bandgap of ~2 eV.

    Illuminated current-voltage characteristics are measured as a function of illumination intensity (Figure 2b). The solar cells show a rectifying characteristic as expected for a *p-n* junction device with a maximum efficiency of 0.016% at 4 suns. Under AM1.5D illumination, the solar cells show a high ideality factor at low illumination conditions prior to decreasing under stronger illumination. This effect may be attributed to a change

in recombination from electron- and hole-limiting at low illumination to being only hole-limited at higher injection conditions, due to the effective mass disparity between the two carrier types.

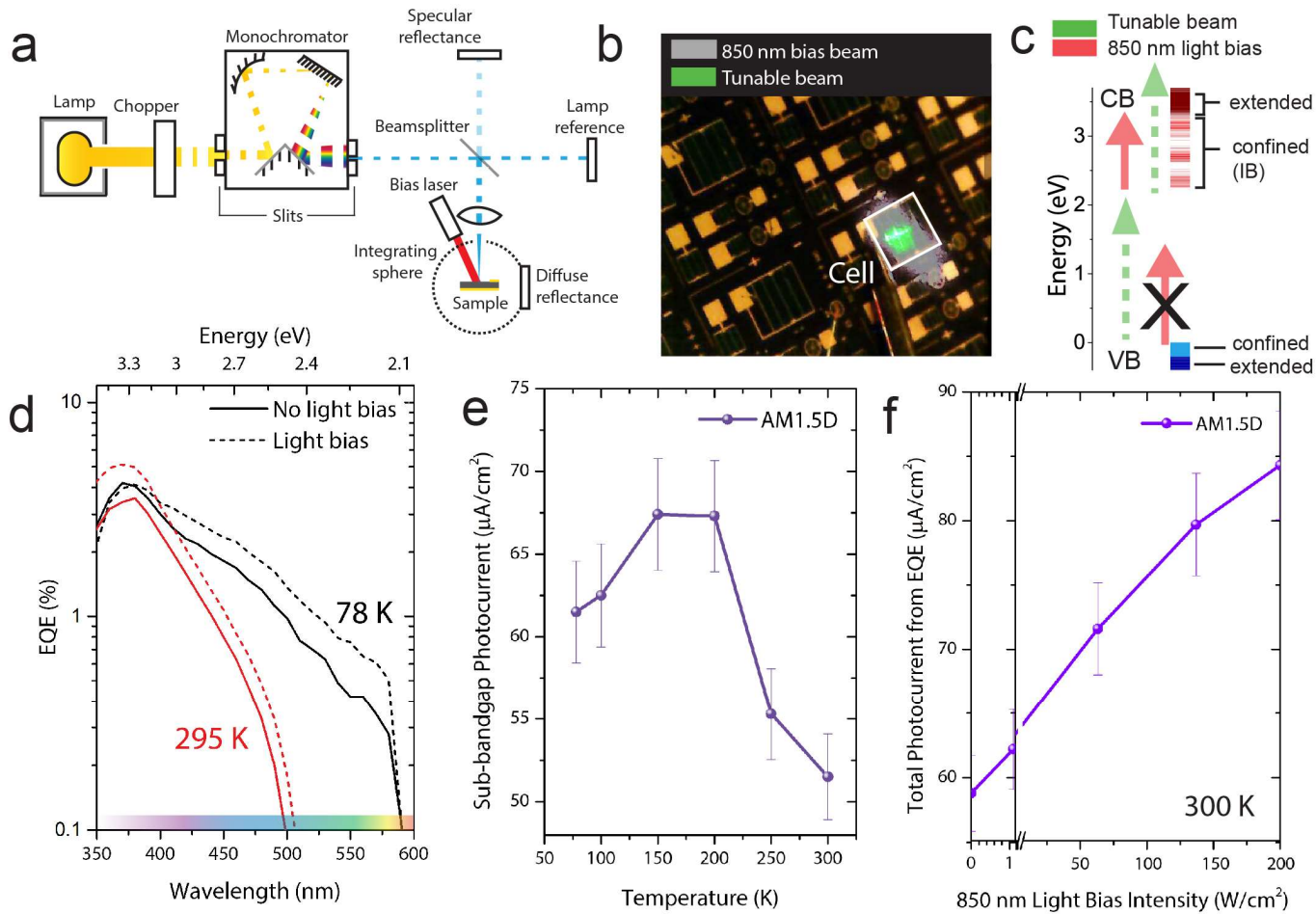

**Figure 3 | Quantum efficiency and photocurrent**. **a** Schematic of the quantum efficiency and reflectivity experimental setup including the laser light bias. **b** Microscope image of the solar cell with the tunable beam (green) and near-infrared light bias beam (grey) spots. **c** Energy levels and transition energies showing the transitions enabled by the tunable beam and the 850 nm light bias. **d** Short circuit quantum efficiency at room temperature and 78 K with and without a 200 W/cm² light bias in a cryostat. **e** Sub-bandgap (>370 nm) photocurrent as a function of temperature without light bias, calculated from quantum efficiency measurements. Lines are guides to the eye. Error bars indicate estimated errors in photocurrent (±5%). **f** Calculated short-circuit current from the quantum efficiency under AM1.5D illumination as a function of light bias intensity at 295 K without a cryostat.

**Simulated maximum absorptance**

We anticipated the quantum efficiency of these devices to be limited by the quantum dot sub-bandgap absorption. We constructed a drift-diffusion model of an InGaN/GaN quantum well device to estimate an upper limit to the sub-bandgap absorptance from the quantum dot region with ideal miniband transport (Figure 2c), detailed in Sec. 6 of the Supplementary Information. This model contrasts with our decoupled dots in the fabricated nanowires, which likely exhibit less effective carrier collection than the miniband model. Nevertheless, the model estimates the effects of incomplete absorption on our device performance. The large diameter of the quantum dots leads to a quantum-well-like density of states in the quantum dots. The simulated absorptance for a quantum dot device is shown in Figure 2c. The model treats InGaN/GaN multiple quantum wells with the same thicknesses as the experimental dots; the 10-layer configuration corresponds most closely to our experimental devices.

The simulated device absorbs 20-30% of the incident light at wavelengths from 500 to 620 nm. Full devices will require nearly 50-100 quantum dots or light trapping enhancements to achieve sufficient absorption. We estimate an effective bulk absorption coefficient for the valence band to intermediate band process from the slope of the absorptance vs. number of quantum dots in Figure 2c, giving a value of approximately $5\times10^5$ cm$^{-1}$ in a broad range from 500-600 nm. We can use this value to estimate the hole figure of merit $v_h$ to be ~100 (assuming $\tau_h$=1 ns and $v_h$ = 10 cm$^2$/V·s), which is compatible with high efficiency devices with active regions of length ~500 nm, corresponding to 100 quantum dots per nanowire[24],which is among the highest predicted values of $v$ for any material[23].

**Two-photon photocurrent**

The hallmark of IBSC operation is current produced by sequential two-photon absorption with energies below the bulk bandgap[12,13,18].We confirm the existence of a quantum-dot mediated sequential two-photon absorption process by measuring the quantum efficiency with a chopped tunable wavelength beam and an 850 nm bias light, as described in Figures 3a and 3b. While the tunable beam can drive both valence band to intermediate band and intermediate band to conduction band transitions, the light bias can only excite the intermediate band to conduction band transitions (Figure 3c). In order to be sensitive to the valence band to intermediate band transition, we use a strong light bias, sufficient to promote most carriers in the intermediate band to the conduction band. The experiment is designed to be limited by the valence band to intermediate band absorption process, which requires weak monochromatic, tunable beam at wavelengths less than 600 nm. The infrared bias photons can only be absorbed if intermediate band states are populated after prior absorption of shorter wavelength light. As shown in Figure 3d, the short-circuit quantum efficiency is maximized at 370 nm, then gradually decreases, yet persists, at wavelengths longer than the GaN band gap even without the light bias. We identify two possibilities for this last effect. The first possibility is the production of a two-photon photocurrent by a single monochromatic beam. The second possibility is current arising from leakage via defect or surface states in a single-photon absorption process. We observe that the photocurrent increases linearly with tunable beam intensity, and therefore deem a two-photon photocurrent process to be unlikely without light bias. The low tunable beam intensity rules out both a significant Auger excitation process and a photofilled intermediate band limiting two-photon process. The subgap photocurrent without light bias is therefore attributed to single-photon absorption leakage pathways. A similar subgap photocurrent is also observed in InGaN quantum well intermediate solar cells on sapphire. [18] With a strong light bias, the photocurrent increases by 44% and 19% at room temperature and 78 K, respectively. At 78 K, this photocurrent increase is predominantly sub-bandgap. Previous observations of sequential two-photon absorption have seen a two-photon signal that is orders of magnitude smaller than the one-photon subgap signal[12,13,18]. The observations in Figure 3d are the largest sequential two-photon signal of which we are aware.

The increase in photocurrent upon the introduction of the light bias is constrained by the population of already-excited dots, which is expected to be small and biased toward the top few quantum dots of the nanowires. These electrons in the intermediate band are then excited to the conduction band by the strong bias light. Without the bias light, many of those electrons fall back to the valence band before they can be promoted to the conduction band. With a strong light bias, the total possible photocurrent gain is limited by the valence band to intermediate band absorption, by construction of this experiment. Figure 3d shows that shorter wavelengths have a much higher contribution to the subgap photocurrent with and without the light bias, which we expect for two reasons. Firstly, the absorption is stronger at short wavelengths. Secondly, their absorption occurs near the top of the device. The much higher electron mobility over the hole mobility causes carrier extraction to be significantly more effective near the top of the nanowire where the holes have higher collection efficiency. At room temperature, the light bias induces a strong photocurrent increase at room temperature below 375 nm that does not appear at 78 K. We do not expect the light bias to improve carrier collection for above-gap photons, so the results at 78 K match expectations while the improvement at 295 K is more surprising and is not fully understood, though a similar effect was seen in the other InGaN system studied in Ref. 18, at room temperature. The sequential two-photon experiment has been performed using standard methods and the strong subgap two-photon signal is clear[12,13]. Supplementary Fig. 6b shows temperature-dependent external quantum efficiency increases relative to room temperature without light bias. Quantum efficiency reduces with temperature above 200 K, which is not the trend observed in Figure 3d. The wavelength dependent quantum efficiency as at various temperatures is shown in Supplementary Fig. 6b.

We expect that most of the sequential two-photon photocurrent originates in the top few dots in an uncoupled dot system. A future device can benefit from modulation doping in the quantum dot region to improve hole transport through the quantum dots[27]. Alternatively, reducing the barrier thicknesses to promote miniband conduction through the multiple quantum dot region can also lead to a more homogenous population of carriers in the intermediate band.

We attribute the sub-bandgap photocurrent increase by the light bias to two-step photon absorption but must rule out alternative mechanisms of tunneling and thermionic emission from dot states. We expect a negligible contribution from tunneling since the higher-indium-fraction dots that contribute to the subgap photocurrent have conduction band states approximately 0.9 eV below the conduction band of GaN, leading to negligible tunneling through the 3 nm GaN barriers. The quantum dot *k.p* calculations accounting for the strain and piezoelectric effects in the quantum dots, incorporating indium diffusion, are described in the Supplementary Simulations. To assess the thermionic contribution, we measured the change in photocurrent from wavelengths longer than 370 nm as a function of temperature (Figure 3e), with the photocurrent calculated from the quantum efficiency measurements. The photocurrent is calculated for the AM1.5D spectrum, which is representative of concentrated sunlight. A solar cell driven by thermionic escape would show an increasing subgap photocurrent with temperature. In contrast, we see the sub-bandgap photocurrent decrease with temperature at temperatures above 200 K. We therefore exclude sample heating from the bias as the origin of the increased external quantum efficiency signal from light biasing. The decrease in photocurrent below 150 K is consistent with Figure 2a, which shows *p*-type carrier freeze-out effects. Calculations in the Supplementary Simulations also show that thermionic escape processes are negligible at the temperature range used in the experiments. Supplementary Fig. 6a shows a similar non-thermionic increase in quantum efficiency with a ~0.1 W/cm$^2$ red (635 nm) light bias at room temperature, where no appreciable heating to the cell is possible.

Varying the light-bias intensity gives confirmation that the bias is sufficiently strong to make the valence band to intermediate band transition limit the photocurrent. Separate quantum efficiency measurements were performed at room temperature outside the cryostat, as a function of light-bias intensity. The short circuit current extracted from these quantum efficiency measurements is shown in Figure 3f. With increasing light bias applied up to 200 W/cm$^2$, the sub-bandgap quantum efficiency increases by 44% to 114 µA/cm$^2$, for a total of 77% of the photocurrent being produced from below the bandgap of the host material. The photocurrent in Figure 3f increases only logarithmically with intensity at the higher intensities, which is consistent with the two-photon process being limited by the valence band to intermediate band absorption, as designed. At light bias intensities of 200 W/cm$^2$, the number of bias photons that arrive within the estimated 1 ns hole lifetime is on the order of 100 per nanowire.

Figure 3d also shows single-photon photocurrent without light bias below the bandgap, as in other intermediate band systems[13,18]. This subgap photocurrent is undesirable for IBSC operation, and future devices will need better barriers to ensure carriers are not directly collected from even the first quantum dot.

## Discussion

To realize efficient IBSCs, the absorption in the quantum dots must be increased through light trapping and/or adding more quantum dots. Light trapping is the preferable option due to the lower mobility-lifetime product of InGaN/GaN multilayer quantum dots and can increase the filling of the quantum dot states. While perfectly Lambertian scattering is known to provide a maximum path length enhancement of 4$n^2$, where $n$ is the index of refraction, absorption increase via light trapping has been shown to surpass the 4$n^2$ limit in certain nanostructures[28]. Such light trapping schemes should therefore be applied to IBSCs whenever possible to reduce the absorber thickness and enhance carrier collection. The inter-dot transport characteristics can also be improved through miniband formation by promoting interdot tunneling, especially for holes.

In summary, we show that InGaN/GaN quantum dot-in-nanowire heterostructures on silicon form IBSCs, enabling sub-bandgap current generation on a lower cost platform. We show the strongest relative change in intermediate band quantum efficiency seen to date due to light bias and the first significant intermediate band photocurrent shown on a silicon substrate or with nanowires. Significant sub-bandgap photocurrent enhancement is observed with a near-infrared light bias and does not increase with temperature. These results suggest that wide-bandgap IBSCs can be achieved on a silicon substrate through nanowire geometries. Future investigations are focused on producing optimal InGaN bandgap combinations for the solar spectrum, increasing sub-bandgap absorption in the quantum dots, and reducing non-radiative recombination in the nanowires through optimizations in device design and fabrication.

## Methods

### Microscopy

Scanning electron microscopy is performed using a Zeiss GeminiSEM 500 at 10 kV accelerating voltage under vacuum. Atomic force microscopy is performed using a Bruker Dimension ICON system using ScanAsyst-AIR probes which have a tip radius of 2 nm.

### Growth and Fabrication

The nanowires are grown directly by molecular beam epitaxy in a Veeco Gen II reactor using a radio frequency plasma-assisted nitrogen source. The growth proceeds under nitrogen-rich conditions on an *n++* silicon (111) substrate. Nanowires are grown with average diameter of 89.3 nm at an areal density of 1 x 10$^{10}$ cm$^{-2}$. The bases of the nanowires are 270 nm sections of *n+* GaN with a silicon dopant concentration of 10$^{19}$ cm$^{-3}$. Ten InGaN quantum dots are then grown in an unintentionally doped GaN region. The tops of the nanowires are 300 nm tall *p+* GaN regions with magnesium dopant concentration of 5 x 10$^{19}$ cm$^{-3}$. Polyimide is used to planarize the cell followed by an etching process to expose the tops of the nanowires. The nanowires begin forming with smaller radii and slightly increase in diameter towards the tops of nanowires as they are grown. The growth is halted prior to coalescence of the nanowires. The top contact is formed by an initial Ti/Au mask layer followed by a 30 nm indium tin oxide layer and finished with a Ti/Au contact on top. The back contact is formed by a planar Ti/Au layer. The electric field created in the nominally intrinsic quantum dot region is used to facilitate carrier transport to the quasi-neutral n-type and p-type regions. The nanowires are fabricated into square devices with approximately 1 x 1 mm, 0.5 mm x 0.5 mm, and 0.35 mm x 0.35 mm dimensions as shown in Supplementary Fig. 1.

**Low Temperature Electroluminescence**

The electroluminescence from the nanowire solar cells is measured as a function of temperature inside a Cryo Industries of America liquid nitrogen-cooled cryostat. The cryostat is evacuated to $10^{-3}$ Pa using a two-stage vacuum pump system. The temperature of the cryostat is controlled to an accuracy of ±5 degrees K with a Cryocon temperature controller, which heated the cryostat sample stage to counteract the cooling from the liquid nitrogen. A small amount of rubber cement is used to secure the sample to the stage. The electrical connection to a specific cell is implemented with a needle probe inside the cryostat wired to a Keithley 2430 sourcemeter. The back of the sample is placed in direct contact with the copper sample stage, which is also wired to the sourcemeter. Emitted light is collected and collimated with an $f$=10 cm fused silica convex lens and focused into an iHR 320 spectrometer with a 1200 lines/mm grating blazed at 600 nm. Time-dependent electroluminescence current effects were observed as a function of bias voltage, as shown in Supplementary Fig. 9. With lower forward bias voltages, the time to achieve a steady state current increases. Electroluminescence spectra were acquired after multiple minutes to reduce this effect.

**Current-voltage characteristics**

Illuminated current-voltage characteristics are measured using a four-probe technique to reduce resistance from the wires. A computer controlled Keithley 2430 sourcemeter is used to measure the photocurrent and apply the bias voltage during the sweep. The sweep time is set to approximately 3 seconds from 0 V to 3 V. The samples are mounted on a gold-coated temperature-controlled stage. Illumination is provided by a Newport/Oriel solar simulator with a 1600 W xenon arc lamp with a filter to best produce the AM1.5D spectrum. A 1-sun Si reference cell obtained from Newport (Oriel) with NIST traceable certification is used to determine the appropriate 1 sun light intensity. The variable intensity is achieved through a variety of perforated nickel filters from Spectrolab of known neutral density transmission, which can reduce the incident light intensity of the light to intensities below the unfiltered 16 suns intensity. The spectrum from the Oriel solar simulator is measured with a fibre-coupled ASD FieldSpec spectroradiometer and is shown in Supplementary Fig. 7. The dark current-voltage characteristic demonstrates an on-to-off ratio of 100 and rectify behaviour both shown in Supplementary Fig. 8.

**Quantum Efficiency**
The spot size of the light bias is approximately 1 mm x 1 mm with a primary beam of about 0.5 mm x 0.5 mm with a total power of about 300 mW. The source for the primary beam is a 300 W xenon arc lamp filtered through a monochromator to produce a quasi-monochromatic beam from 300 to 1800 nm with a 5 nm spectral width at half maximum. Room temperature quantum efficiency measurements are performed using a Newport IQE-200 quantum efficiency measurement system coupled to a Merlin lock-in amplifier. The primary monochromatic beam is chopped at a frequency of 87 Hz.
For the low temperature quantum efficiency measurements, the sample is placed on a vertically mounted cryostat cooled with liquid nitrogen, as shown in Supplementary Fig. 5. Sample temperature is measured and maintained with a thermocouple and heater, respectively, both embedded in the copper sample stage. The cryostat is filled with nitrogen and cooled to 78 K. The primary beam is directed onto the sample using a silver mirror. The back contact of the solar cell is electrically connected to a copper cold finger through the mechanical force from a probe needle used to connect to the top contact of the solar cell. The quantum efficiency measurement system is recalibrated to account for the silver mirror and UV-fused silica cryostat window. The continuous wave bias laser beam is incident on the sample at an angle of 10° from the perpendicular axis of the solar cell. A CMOS camera is used to view the overlap of the chopped beam and bias beam on the cell. Beam alignment on the cell is further verified by aligning the electroluminescence emission of the cell to the diffuse reflection of the light bias beam off the solar cell. The quantum efficiency measurements are performed under short-circuit conditions. Higher temperatures than 78 K are achieved by driving current through the heater, causing liquid nitrogen to boil off and a higher temperature near the sample.

**Simulations**

The absorption calculation in Figure 2c is performed with Crosslight APSYS software. The model is based on a standard drift-diffusion semiconductor device coupled to a miniband model for the multiple dot region in the middle of the nanowire. The miniband model is chosen to represent the case that maximizes the transport characteristics in the quantum dot region, so that transport does not limit device performance. Lifetimes and mobilities are set to high values to achieve near-complete carrier collection of sub-bandgap photons. In this quantum well approximation, the maximum absorptance can be calculated as a function of wavelength to approximately assess the fraction of light that could be expected to produce photocurrent, assuming the IB to CB transitions are not limiting. The miniband condition is enforced to maximize transport efficiency in an ideal scenario. No reflection at the top surface is allowed, and a virtual contact is placed just below the $n^+$-GaN base with a reflection condition that corresponds to the refractive index between GaN and silicon. A $p^+$-GaN top emitter thickness of 300 nm, bottom $n^+$-GaN base thickness of 270 nm and 1 to 100 of 3 nm InGaN / 3 nm GaN quantum dots are used, as in our nanowires.

## Acknowledgments

The authors would like to thank the National Science and Engineering Research Council of Canada (NSERC) for the funding of this project under the Photovoltaic Innovation Network (PVIN) Project 11. The authors would also like the thank Ontario Graduate Scholarship (OGS), Canada Foundation for Innovation (CFI), and the Canada Research Chairs program for their support and resources that were essential to make this work possible. Growth and fabrication of the nanowire solar cells were performed in the Microfabrication Facility at McGill University. We would like to acknowledge CMC Microsystems for the provision of products and services that facilitated this research, including Crosslight simulation package and fabrication fund assistance at McGill University.

## Data Availability

Data related to the figures can be found at 10.6084/m9.figshare.12456158. Other data related to this work are available from the authors upon reasonable request.

## Author Contributions

K. H and Z. M. proposed and supervised the research project. The two first authors contributed equally to this research. S. S. performed the growth and fabrication of the nanowires. R.C performed the experimental characterization and simulation of the nanowire devices. R.C. drafted most of the manuscript with help from S. S., J. K, and K. H. L. R. performed the quantum dot $k \cdot p$ calculations. J. K. analyzed and guided the interpretation of the experimental data. All authors have provided comments on the manuscript.

## Competing interests

The authors declare no competing interests.